\documentclass[12pt]{article}
\usepackage{graphics}
\usepackage{graphicx}
\usepackage{amssymb}
\usepackage{amsmath}

\newcommand{\rf}[1]{(\ref{#1})}
\newcommand{\beq}{\begin{equation}}
\newcommand{\eeq}{\end{equation}}
\newcommand{\bea}{\begin{eqnarray}}
\newcommand{\eea}{\end{eqnarray}}

\newcommand{\e}{\mbox{e}}
\renewcommand{\d}{\mbox{d}}

\newcommand{\Lam}{\Lambda}
\renewcommand{\a}{\alpha}
\newcommand{\n}{\nu}

\newcommand{\dph}{\Delta\varphi}

\renewcommand{\th}{\theta}
\newcommand{\om}{\omega}
\newcommand{\del}{\delta}

\newcommand{\Th}{\Theta}
\newcommand{\Tha}{\Th^{(1)}}
\newcommand{\Thb}{\Th^{(2)}}

\newcommand{\oh}{\frac{1}{2}}

\newcommand{\tr}{\mathrm{tr}\,}

\newcommand{\cC}{{\cal C}}

\newcommand{\hx}{{\hat{x}}}
\newcommand{\hy}{{\hat{y}}}

\newcommand{\bx}{{\bar{x}}}

\newcommand{\SL}{\sqrt{\Lam}}

\begin{document}

\begin{center}
\vspace{24pt}
{ \large \bf Universality of 2d causal dynamical triangulations}

\vspace{30pt}

{\sl J. Ambj\o rn}$\,^{a,b}$
and 
{\sl A. Ipsen}$\,^{a}$

\vspace{48pt}
{\footnotesize

$^a$~The Niels Bohr Institute, Copenhagen University\\
Blegdamsvej 17, DK-2100 Copenhagen \O , Denmark.\\
{email: ambjorn@nbi.dk, acipsen@gmail.com}\\

\vspace{10pt}

$^b$~Institute for Mathematics, Astrophysics and Particle Physics (IMAPP)\\ 
Radbaud University Nijmegen, Heyendaalseweg 135,
6525 AJ, Nijmegen, The Netherlands.

}
\vspace{96pt}
\end{center}


\begin{center}
{\bf Abstract}
\end{center}

The formalism of Causal Dynamical Triangulations (CDT)  attempts 
to provide a non-perturbative regularization 
of quantum gravity, viewed as an ordinary quantum field theory. 
In two dimensions one can solve the lattice theory analytically
and the continuum limit is universal, not depending on the 
details of the lattice regularization.

\vspace{12pt}
\noindent

\vspace{24pt}

\newpage

\section{Introduction}\label{intro}

Two-dimensional quantum gravity has been a fruitful laboratory 
for studying aspects of string theory as well as quantum gravity.
One somewhat surprising aspect of Euclidean two-dimension quantum 
gravity coupled to matter in the form of a conformal field theory,
is that the regularized lattice theory, using the so-called dynamical
triangulations (DT), can be solved analytically. 
The details of the DT regularization are unimportant 
for the continuum limit. In fact it has been a wonderful example of 
universality in the Wilsonian sense, the critical surface 
where the continuum limit can be taken being of finite co-dimension 
in an infinite dimensional coupling constant space (see e.g.\ \cite{book}
for a review).
The lattice regularization known as causal dynamical triangulations (CDT)
uses a subset of the triangulations used in DT  \cite{al,ajl}. The original 
idea was to consider a path integral where spacetime histories 
before rotating to Euclidean signature were locally causal, i.e.
had non-degenerate light cones (see  
\cite{review} for a review of the CDT approach also in higher dimensions
than two). 
In two dimensions, which is the only 
case we will consider here, the precise relation between the CDT triangulations
and the DT triangulations was described in \cite{ackl}.

There is good evidence of universality of the CDT scaling limit, although 
one does not have the same comprehensive evidence as for the DT case. First, 
a related model, in a certain way more general, the so-called string-bit 
model \cite{durhuus}, led to the same scaling limit. Further it 
was shown in \cite{dgk1} that one could add dimers on the ``spatial''
CDT links without changing the universality class. Thus it was somewhat 
surprising that adding further ``dressing'', but only 
along the spatial links, seemingly led to new continuum models, depending on a 
continuous parameter $\beta$ (to be defined below) \cite{dgk2}. 
The purpose of this letter
is to show that also for this general set of models one obtains indeed 
the standard CDT scaling limit. 

\section{Defining the model}\label{definition}

\begin{figure}[t]
\centering
\includegraphics[width=0.9\textwidth]{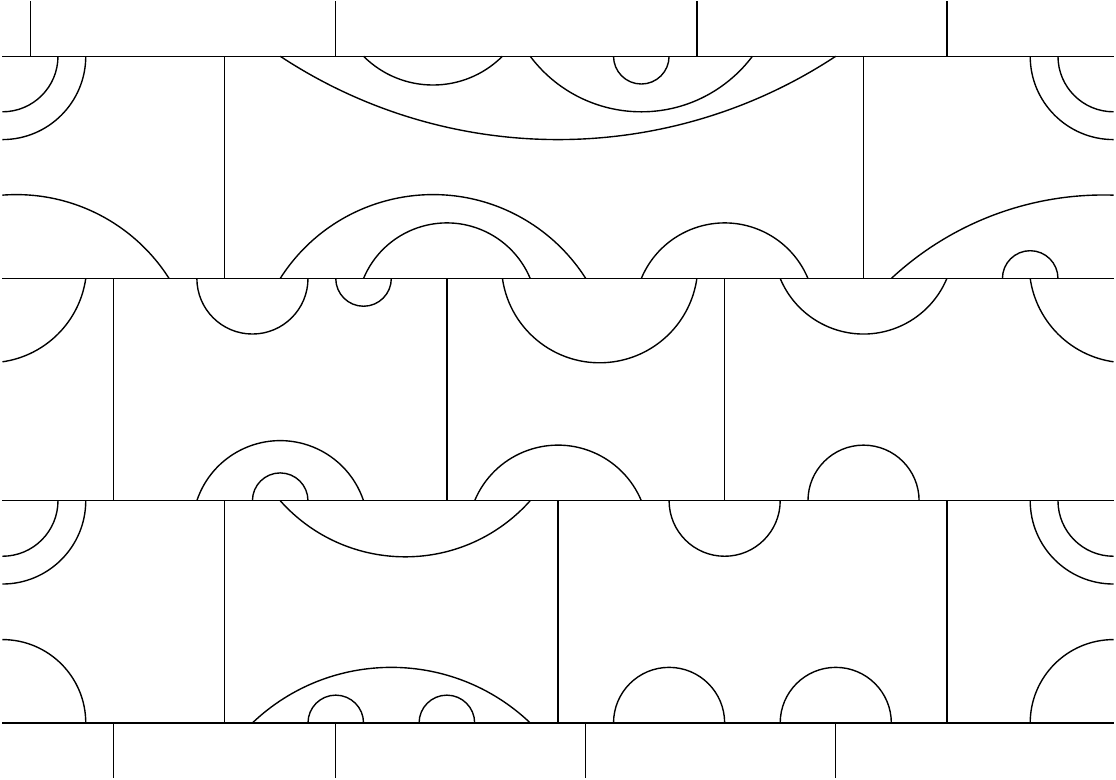}
\caption[fig1]{Modified CDT configuration, dual graph.}
\label{fig1}
\end{figure}

The modified CDT model (not to be mistaken for what has later 
been called ``generalized CDT'' \cite{generalCDT}) is most easily defined using
a lattice dual to the triangulation, i.e.\ a $\phi^3$ graph with 
a ``time'' foliation . Fig.\ \ref{fig1} shows the dual CDT lattice 
and its generalization. In this dual picture  each vertex represents
a triangle in the ``original'' triangulation and each polygon represents 
a vertex, the order of which is equal the number of sides in the polygon.

In the modified model one allows a dressing of the horizontal links 
between two vertical links by rainbow diagrams.

Three coupling constants are assigned to the model: to each vertex 
one associates a coupling constant $g$, to a vertex with an incident 
vertical  link an additional  coupling constant $h$, and finally     
to each vertex with an incident rainbow link a coupling constant $\th$.
The parameter 
\beq\label{j0}
\beta = \frac{\th}{h}
\eeq
governs the density of rainbow links compared to the number 
of vertical links, i.e. ``time-like'' links in the original CDT-like 
$\phi^3$-graph. In this article we will only consider $0\leq \beta<1$,
which is the range leading to CDT-like theories \cite{dgk2}.

As shown in \cite{dgk2} one can define and calculate a transfer matrix 
for this model. The result is 
\beq\label{j1}
\Th_{ij} = \sum_k \Thb_{ik}\Tha_{kj}
\eeq
where the index $j$ refers to the number of incoming half-lines which 
is incident from below on the horizontal line at time $t$ and index $k$ 
refers to the number of half-lines leaving the horizontal line at time $t$.
Index $k$ plays the same role as index $j$, only at time-slice $t+1$. In this 
way $\Thb_{ik}$ connects outgoing vertical half-lines  at $t$ to incoming  
half-lines at $t+1$ and $\Th_{ij}$ incoming half-lines at $t$ to incoming 
half-lines at $t+1$.

$\Tha$ is the CDT transfer matrix, already 
discussed in \cite{al} and analyzed in detail in \cite{dgk1}. 
If $\th=0$ and $h=1$ there are  
no rainbow lines and $\Thb$ becomes the identity matrix and $\Th$ 
also the CDT transfer matrix.    
 
It is convenient to work with the discrete Laplace transforms of 
$\Th$, $\Tha$ and $\Thb$. To simplify the expressions somewhat 
we make the following redefinitions compared to \cite{dgk2}:
\beq\label{redef}
\Tha_{ij} \to (2g)^{-i-j} \Tha_{ij},~~~~\Thb_{ij} \to (2g)^{i+j} \Thb_{ij}.
\eeq  
The explicit expressions are then:
\bea
\Tha(x,y) &=& \sum_{ij} x^iy^j \Tha_{ij} = \frac{1}{1-\oh x-\oh y} \label{j3}\\
 \Thb(x,y) &=& \frac{C(\hx^2) C(\hy^2)}{(1-\hx^2C(\hx^2))
(1-\hy^2C(\hy^2))(1-\beta^{-2}\hx\hy\,C(\hx^2)C(\hy^2))} \label{j4}\\
\Th(x,y) &=& \oint_{\cC} \frac{d\om}{2\pi i\om} \; \Tha(x,\om^{-1})\Thb(\om,y),
\label{j5}
\eea 
where the contour encloses cuts and poles and where
\beq\label{j6}   
\hx=2g\th x,~~~\hy=2g\th y,~~~~C(z) = \frac{1-\sqrt{1-4z}}{2z}.
\eeq

Integrating over the simple pole of $\Tha$ one obtains
\beq\label{j6a}
\Th(x,y)= \frac{1}{1-\oh x}\frac{C(\bx^2) C(\hy^2)}{(1-\bx^2C(\bx^2))
(1-\hy^2C(\hy^2))}\frac{1}{1-\beta^{-2}\bx \hy C(\bx^2)C(\hy^2)},
\eeq
where 
\beq\label{j6b}
\bx= \frac{2g \th}{2-x}
\eeq

The partition function with open horizontal boundaries after $t$ time steps 
is\footnote{The same continuum limit is obtained by setting 
$Z(l,k;t) = \big( \Theta^t\big)_{kl}$.}
\beq\label{j7}
Z(l,k;t) = \Big( (\Tha (\Thb\Tha)^t\Big)_{kl},
\eeq
and the (discrete) Laplace transformed function is denoted $Z(x,y)$
\beq\label{j8}
Z(x,y;t) = \sum_{l,k} x^ly^k Z(l,k;t).
\eeq
The partition function after $t$ time steps with periodic boundary 
conditions in the time direction is
\beq\label{j9}
Z(t) =\tr (\Th^t).
\eeq

\section{The continuum limit using the transfer matrix}\label{transfer}

As shown in \cite{dgk2} the partition function $Z(t)$ has a singularity at 
\beq\label{j10}
\xi_c= 2g\th\left(\beta+\frac{1}{\beta}\right)=1.
\eeq
We want to take to continuum limit by approaching this singularity.
This is done in the following way \cite{dgk2}:
\beq\label{j11}
\xi\equiv 2g \th \left(\beta+\frac{1}{\beta}\right) = 
1-\oh a^2 \Lam \,\left(\frac{1-\beta^2}{1+\beta^2}\right)^2.
\eeq
The interpretation is that $a$ is the lattice spacing,
i.e.\ the link length in the triangulation, and $\Lam$
the cosmological constant, such that the average number of 
triangles is proportional to $1/(\Lam a^2)$. Thus the average 
``continuum'' area is proportional to $1/\Lam$.

Until now $t$ has denoted the integer number of time steps 
in the triangulation. We are interested in a limit where
we have a finite continuum time $T$ scaling as
\beq\label{j12}
T = t a,
\eeq
where $a$ is the lattice spacing defined by \rf{j11}.
We can then write 
\beq\label{j13}
Z(T) = \tr \Th^t = \tr e^{-T H},~~~\Th = \e^{-a H}.
\eeq
Thus an expansion of $\Th$ to lowest order in $a$
should allow us to determine $H$.

If the continuum area is proportional to $1/\Lam$ we expect the 
continuum length of a time slice to be proportional to $1/(\Lam T)$. 
Thus we expect a scaling $L \propto l \,a$ where $l$ is the number 
of space-like links. We can also enforce this on the boundaries:
\beq\label{j13a}
Z(l,k;t) \to Z(L_0,L_T;T).
\eeq  
The discrete Laplace transform of $Z(x,y;t)$ has poles in $x,y$ and 
it is at these poles one extracts the continuum function $Z(L_0,L_T;T)$.
These poles are at $x_c=y_c=1$
for $a \to 0$. The terms $x^l$ and $y^k$ in \rf{j8} can then 
be given an interpretation as the part  of the action 
coming from a continuum boundary cosmological term proportional to $X$
if we scale:
\beq\label{j13b}
x=1-aX \,\left(\frac{1-\beta^2}{1+\beta^2}\right)^2,~~~~~
L= a \,l \left(\frac{1-\beta^2}{1+\beta^2}\right)^2,
\eeq
and thus
\beq\label{j13bb}
x^l\to \e^{-L X }~~~{\rm for}~~~a \to 0.
\eeq 
With this scaling we obtain a relation similar to \rf{j13a}, going 
from the discretized expression to the continuum expression:
\beq\label{j13c}
Z(x,y,t)\to Z(X,Y;T),
\eeq
where the continuum analogue of \rf{j8} reads
\beq\label{j13d}
Z(X,Y;T) = \int_0^\infty \d L_0 \d L_T \; \e^{-L_0X -L_TY} Z(L_0,L_T;T).
\eeq 
We will return to \rf{j13a} and \rf{j13c} in the next section. 

We now extract $H$ from $\Th= e^{-aH}$. It is convenient 
to use the Laplace transform \rf{j5} of $\Th$. Expanding in $a$ 
we obtain \cite{dgk2}:
\beq\label{j14}
 \Big((1-a H +O(a^2)) \psi\Big)(x)= 
\oh\frac{1-\beta^2}{1+\beta^2} \oint \frac{d\om}{2\pi i \om}\; 
\Th\Big(x,\frac{1}{\om}\Big) \, \psi(\om).
\eeq
Here $\psi(\om)$ is the discrete Laplace transform of 
a function $\psi(l)$:
\beq\label{j14a}
\psi(\om) = \sum_l \om^l \psi(l). 
\eeq
The function $\Th(x,1/\om)$ has a pole in $\om$ at 1 for $a\to 0$ and it 
has a branch cut located at $\om \in 
[-\om_*,\om_*]$, where 
\beq\label{j19} 
\omega_{*} = 2\left(\beta+\frac{1}{\beta}\right)^{-1}+O(a) < 1
~~~\mbox{for $a$ sufficiently small}.
\eeq   
We can deform the contour to be a small circle around one and 
an integration along the branch cut. The integration around
$\om =1$ allows us to use the expansion \rf{j13b} for $x$ and $\om$,
and we obtain
\beq\label{20}
 \oint\frac{d Z}{2\pi i}\left[
  \frac{1}{Z-X}+\frac{a}{(Z-X)^2}\left(
  \Lambda+\frac{\beta^2X^2-(1+3\beta^2)XZ+\beta^2Z^2}{1+\beta^2}
  \right)
  \right]\psi(Z) + O(a^2),
\eeq 
Performing the integration (and ignoring the contribution from the cut)
we can identify $H$ as 
\beq\label{j18a}
H(X)= (X^2-\Lambda)\frac{\partial}{\partial X}+X,
\eeq
and by an inverse Laplace transformation
\beq\label{j18b}
 H(L)=-L\frac{\partial^2}{\partial L^2}-
\frac{\partial}{\partial L} + \Lambda L.
\eeq
This is precisely the ordinary CDT Hamiltonian, the only 
difference is that in order to obtain it in this form we
had to perform a dressing (or renormalization) of the 
continuum boundary cosmological constant from a value $X$,
corresponding to $\beta=0$ to the $\beta$ dependent value given in 
\rf{j13b}. This renormalization of $X$ and a similar 
renormalization of the coupling  cosmological coupling 
constant $\Lam$ in \rf{j11} is all that is needed to include
the effects of the rainbow diagrams.

The contribution from the cut can be written as  
\beq\label{j20}
\tilde\psi(x) = \int_{-\omega_*}^{\omega_*}d\omega f(x,\omega)\psi(\omega),
\eeq
where $f(x,\om)$ is integrable in $[-\om_*,\om_*]$ and  
$\tilde{\psi}(x)$ analytic in the neighborhood of $1$ and finite 
when $a\to 0$. We cannot view such a  function as the 
Laplace transform of any function $\psi(\SL L)$ depending 
on the continuum length $L >0$, the reason being that the inverse 
Laplace transformation from \rf{j18a} to \rf{j18b} gives
\beq\label{j21}
\int_{i\infty+c}^{i\infty +c} \frac{dX}{2\pi i} 
\;\e^{XL} \tilde{\psi}(1-aX)= 
\del (L) \tilde{\psi}(1)-a \del'(L) \tilde{\psi}'(1) + \cdots 
+O(a^n).
\eeq
Thus we do not associate any continuum physics with the analytic 
function $\tilde{\psi}(x)$ defined by \rf{j20}
\footnote{Of course a function like $\tilde{\psi}(\om)$ would also 
not contribute to continuum physics if inserted in \rf{20}.
The part of a function $\psi(\om)$ defined as in \rf{j14a}
which {\it does} contribute to 
continuum physics in \rf{20} is the part which has a continuum
Laplace transform, i.e. the part where $\psi(l)$ in \rf{j14a} 
has the form 
$\psi(\sqrt{\xi-\xi_c} \, l) \to \psi(\SL L)$. Since $\sqrt{\xi-\xi_c} 
\propto a \SL$ it can at most be the tail at infinite $l$ which 
contributes to continuum physics for a given 
$\psi(\om) = \sum_l \om^l \psi(l)$.}.

\section{The Schwinger representation and the continuum}
\label{schwinger} 

In \cite{dgk2} the modified CDT Hamiltonian was not derived 
using the transfer matrix as described above, but rather 
a so-called Schwinger representation of $Z(x,y;t)$. We now 
show that this method also leads to \rf{j18b}, i.e.\ the ordinary 
CDT Hamiltonian.  

The starting point is the following representation of $Z(x,y;t)$
(\cite{dgk2}, formula (5.19)):
\beq\label{j30}
  Z(x,y;t) = \prod_{s=0}^t\left(\int_0^\infty d\alpha_s e^{-\alpha_s}\right)
  e^{\frac{1}{2}(\alpha_0 x+\alpha_t y)}\prod_{r=0}^{t-1}
  \phi_\beta(g\theta \alpha_r,g\theta\alpha_{r+1})
\eeq
where 
\beq\label{j31}
\phi_\beta(x,y) = \sum_{k \geq 0} I_k(2x)I_k(2y)/\beta^{2k}.
\eeq
$x$ and $y$ only appears in the exponential function and we can write
\beq\label{j32}
  Z(x,y;t) = \int_0^\infty d\a_0 \int_0^{\infty} d\a_t 
  \,\e^{-\oh(1-x)\a_0-\oh(1-y)\a_t} F(\a_0,\a_t;t),
\eeq
where 
\beq\label{j33}
F(\a_0,\a_t;t) = 
\left(\prod_{s=1}^{t-1}\int_0^\infty d\alpha_s\right) 
  \prod_{r=0}^{t-1}
 \e^{-(\a_r+\a_{r+1})/2} \phi_\beta(g\theta \alpha_r,g\theta\alpha_{r+1}).
\eeq 
Since $1-x \propto a X$ and $1-y\propto a Y$, \rf{j32} 
states that in the limit where $a\to 0$ and thus $Z(x,y;t) \to Z(X,Y;T)$,
$Z(X,Y;T)$ is the Laplace transform of $F(\a_0,\a_t;t)$, $t=T/a$. Thus,
in accordance with \rf{j13d} we have
\beq\label{j34}
F(\a_0,\a_t;t)\propto Z(L_0,L_T;T),
\eeq
where
\beq\label{j34a}
L_0= \oh a\a_0\left(\frac{1-\beta^2}{1+\beta^2}\right)^2,
~~L_T=\oh a \a_t\left(\frac{1-\beta^2}{1+\beta^2}\right)^2,~~ 
a\,t=T.
\eeq
If we change variables from $\a_s$ to $\varphi_s$,
\beq\label{j35}
  \alpha_s = \frac{\varphi_s^2}{a}\left(\frac{1+\beta^2}{1-\beta^2}\right)^2,
\eeq
we obtain
\beq\label{36}
  Z(L_0,L_T;T) \propto\frac{1}{\sqrt{\varphi_0\varphi_t}} 
  \int_0^\infty \prod_{s=1}^{t-1}\d\varphi_s\; 
  \prod_{r=0}^{t-1}
  \frac{\sqrt{\varphi_r\varphi_{r+1}}}{a}
  \frac{1+\beta^2}{1-\beta^2}
  \e^{-\frac {\alpha_r + \alpha_{r+1}} 2}
  \phi_\beta(g\theta \alpha_r,g\theta\alpha_{r+1}).
\eeq
The right hand side can be interpreted as a (quantum mechanical) 
path integral, i.e.\  
\beq\label{j37}
  \sqrt{\varphi_0\varphi_t}\;Z(L_0,L_T;T) 
\propto \langle \varphi_0|e^{-TH}|\varphi_t\rangle
\eeq
for some Hamiltonian $H$. We will now proceed to determine $H$.

Following \cite{dgk2}  we use the notation 
\beq\label{j38}
  \e^{-\frac{\alpha_0 + \alpha_{1}}{2}}
  \phi_\beta(g\theta \alpha_0,g\theta\alpha_{1})
  \sim U_\beta(\alpha_0,\alpha_1)\;\e^{-S_\beta(\alpha_0,\alpha_1)}.
\eeq
According to \cite{dgk2}  
\beq\label{j39}
  S_\beta(\alpha_0,\alpha_1) = \frac 1 2(\alpha_0+\alpha_1)
  -2g\theta\sqrt{(\alpha_0+\beta^2\alpha_1)(\alpha_0+\beta^{-2}\alpha_1)}
\eeq
and 
\begin{multline}
  U_\beta(\alpha_0,\alpha_1) = \frac{1}{\sqrt{4\pi g\theta}}
  \frac 1 {((\alpha_0+\beta^2\alpha_1)(\alpha_0+\beta^{-2}\alpha_1))^{1/4}}\\
  \times \left(1
  +\frac 1 {16g\theta\sqrt{(\alpha_0+\beta^2\alpha_1)(\alpha_0+\beta^{-2}\alpha_1)}}
  +\cdots\right).\label{j40}
\end{multline}

We now expand in $a$, with
\beq\label{j40a}
  \dph = \varphi_1-\varphi_0
\eeq
counted as being of order $\sqrt a$ as one has to do in a path 
integral (here we differ from \cite{dgk2}):
\beq\label{j41}
  S_\beta(\alpha_0,\alpha_1) = \frac{\dph^2}{2a}
  -\frac{\beta^2}{(1+\beta^2)^2}\frac{\dph^4}{2a\varphi_0^2}
  +\frac{a\Lambda}{2}\varphi_0^2 + O(a^{3/2}).
\eeq
We see that we get a standard kinetic term, justifying $\dph \propto \sqrt a$.
(Note that the $\dph^4$ term is not present in \cite{dgk2}).

Similarly, we find
\beq\label{j44}
  \frac{\sqrt{\varphi_r\varphi_{r+1}}}{a}
  \frac{1+\beta^2}{1-\beta^2}U_\beta(\alpha_0,\alpha_1) =
  \frac{1}{\sqrt{2\pi a}}\left(1+\frac{a}{8\varphi_0^2}
  -\frac{\beta^2}{(1+\beta^2)^2}\frac{\dph^2}{\varphi_0^2}+O(a^{3/2})\right).
\eeq
(We note that the $\dph^2$ term is not present in \cite{dgk2}.)

The Hamilton is finally determined by integrating against a trial state:
\begin{multline}\label{j46}
  ((1-aH)\psi)(\varphi_0) =
  \int_0^\infty\frac{\d \varphi_1}{\sqrt{2\pi a}}e^{-\frac {\dph^2} {2a}}
  \biggl[1
  +\left(\frac{1-\beta^2}{1+\beta^2}\right)^2\frac{a}{8\varphi_0^2}
  -\frac{\beta^2}{(1+\beta^2)^2}\frac{\dph^2}{\varphi_0^2}\\
  +\frac{\beta^2}{(1+\beta^2)^2}\frac{\dph^4}{2a\varphi_0^2}
  -\frac{a\Lambda}{2}
\varphi_0^2 \biggr]
  \left[1+\dph\frac{\partial}{\partial\varphi}
  +\frac {\dph^2} 2 \frac{\partial^2}{\partial\varphi^2}\right]\psi(\varphi_1).
\end{multline}
Carrying out the Gaussian integral, we obtain
\beq\label{j47}
  H = -\frac 1 2\frac{\partial^2}{\partial\varphi^2}
  +\frac{\Lambda}{2}\varphi^2
  -\frac{1}{8\varphi^2}.
\eeq
This is precisely the CDT Hamiltonian when changing back to the $L$ 
variable.
  
\section{Critical arches}

In principle a new behavior could be possible for 
$\beta \to 1$ from below, since in this case the 
rescaling  of lengths and boundary cosmological 
constants, as defined by eqs.\ \rf{j13b}, diverges
and it is precisely the limit where the cut will merge with 
the pole in the expression \rf{j6a} for $\Th$.
Let us investigate this case by assuming
\beq\label{j60}
\beta = 1 -a^{\eta} B,
\eeq
where $B$ is a new physical constant with mass dimension $\eta$.
To understand the analytic structure of $\Th$ for $a\to 0$, i.e.\
$\beta \to 1$ from below, we expand  the argument of the square root 
related to the Catalan number in the expression for $\Th$:
\beq\label{j61}
\sqrt{1-4\hat{x}^2} = a^{\eta}B(1+a X +\oh a^2 (\Lam-X^2) + O(a^\eta B)+ O(a^3))
\eeq
From this expression it is clear that that the cut has 
disappeared  from the expression even though it hits the pole 
when expressed in terms of unrenormalized variables.  
To find the Hamiltonian we use the same approach as in Sec.\ \ref{transfer},
eqs.\ \rf{j14} and \rf{20}
and write
\beq\label{j70}
\tilde{\psi}(x) = (1-a^\n H +\cdots)\psi(x) := \frac{a^\eta B}{2}
\oint\frac{\d \om}{2\pi i \om} \;\Th(x,\frac{1}{\om}) \psi(\om),
\eeq
where $\n$ is determined by the expansion, We find:
\beq\label{j71}
\tilde{\psi}(X) = \oint \frac{\d Z}{2\pi i} \left[ 
\frac{1-a^\eta {B}/{2}}{Z-X} + a\;\frac{\Lam +\oh (X^2-4XZ+Z^2)}{(Z-X)^2} \right]
\psi(Z).
\eeq
Thus, if $\eta >1$ we obtain the same results as before 
(eq.\ \rf{20} with $\beta=1$) and if $\eta<1 $
we obtain a trivial Hamiltonian. $\eta =1$ just adds  the 
positive constant $B/2$ to the CDT Hamiltonian \rf{j18a}.
So far we have ignored the contributions from the cut. However,
arguments like the ones used in Sec.\ \ref{transfer} show that 
the cut will not contribute in the scaling limit.

\section{Discussion}

We have shown that the CDT scaling limit is quite universal
and independent of details of the lattice regularization, as 
long as we maintain a reasonable ``memory'' of the underlying
assumed time foliation. Dressing the spatial slices with 
a few outgrowths should not alter the scaling limit 
and this is indeed what we have proven to be the case.
Potentially there could have been a different behavior 
in the limit $\beta \to 1$ where the rainbow diagrams 
become critical, but explicit calculations showed that 
it was not the case. The CDT model provides us with
a regularized of a theory of fluctuating spacetime which is invariant
under spatial diffeomorphisms and which allows for a time
foliation. The simplest such continuum model is a Ho\v rava-Lifshitz
gravity model in two-dimensions where we only keep terms with 
at most second order derivatives of the metric, and one 
can indeed show that such a model has a classical CDT Hamiltonian 
which when quantized is compatible with the $H(L)$ considered in
this paper \cite{als}.

\subsection*{Acknowledgments}
The authors thank C.F. Kristjansen for helpful discussions.
They also acknowledge support from the ERC-Advance grant 291092,
``Exploring the Quantum Universe'' (EQU). JA acknowledges support 
of FNU, the Free Danish Research Council, from the grant 
``quantum gravity and the role of black holes''.


\begin{thebibliography}{99}


\bibitem{book}
  J.~Ambjorn, B.~Durhuus and T.~Jonsson,
  Cambridge, UK: Univ. Pr., 1997. (Cambridge Monographs in Mathematical Physics). 363 p


\bibitem{al}
  J.~Ambjorn and R.~Loll,
  Nucl.\ Phys.\ B {\bf 536} (1998) 407
  [hep-th/9805108].

\bibitem{ajl}
  J.~Ambjorn, J.~Jurkiewicz and R.~Loll,
  Nucl.\ Phys.\ B {\bf 610} (2001) 347
  [hep-th/0105267].



\bibitem{review}
  J.~Ambjorn, A.~Goerlich, J.~Jurkiewicz and R.~Loll,
Physics Reports, {\bf 519} (2012) 127-210
  arXiv:1203.3591 [hep-th].




\bibitem{ackl}
  J.~Ambjorn, J.~Correia, C.~Kristjansen and R.~Loll,
  Phys.\ Lett.\ B {\bf 475} (2000) 24
  [hep-th/9912267].

\bibitem{durhuus}
  B.~Durhuus and C.~W.~H.~Lee,
  Nucl.\ Phys.\ B {\bf 623} (2002) 201
  [hep-th/0108149].



\bibitem{dgk1}
  P.~Di Francesco, E.~Guitter and C.~Kristjansen,
  Nucl.\ Phys.\ B {\bf 567} (2000) 515
  [hep-th/9907084].



\bibitem{dgk2}
  P.~Di Francesco, E.~Guitter and C.~Kristjansen,
  Nucl.\ Phys.\ B {\bf 608} (2001) 485
  [hep-th/0010259].

\bibitem{generalCDT}
  J.~Ambjorn, R.~Loll, Y.~Watabiki, W.~Westra and S.~Zohren,
  Phys.\ Lett.\ B {\bf 670} (2008) 224
  [arXiv:0810.2408 [hep-th]];
  Phys.\ Lett.\ B {\bf 665} (2008) 252
  [arXiv:0804.0252 [hep-th]];
  JHEP {\bf 0805} (2008) 032
  [arXiv:0802.0719 [hep-th]].\\
  J.~Ambjorn, R.~Loll, W.~Westra and S.~Zohren,
  JHEP {\bf 0712} (2007) 017
  [arXiv:0709.2784 [gr-qc]].

\bibitem{als}
J. Ambjorn, L.Glaser, Y. Sato and Y. Watabiki, to appear.
 
\end{thebibliography}
\end{document}